# Prism-array lenses for energy filtering in medical x-ray imaging

Erik Fredenberg,[a] Björn Cederström,[a] Carolina Ribbing,[b] and Mats Danielsson[a]

[a]Department of Physics, Royal Institute of Technology, AlbaNova, 106 91 Stockholm, Sweden;

[b]The Ångström Laboratory, Uppsala University, 751 21 Uppsala, Sweden;

## ABSTRACT

Conventional energy filters for x-ray imaging are based on absorbing materials which attenuate low energy photons, sometimes combined with an absorption edge, thus also discriminating towards photons of higher energies. These filters are fairly inefficient, in particular for photons of higher energies, and other methods for achieving a narrower bandwidth have been proposed. Such methods include various types of monochromators, based on for instance mosaic crystals or refractive multi-prism x-ray lenses (MPL's). Prism-array lenses (PAL's) are similar to MPL's, but are shorter, have larger apertures, and higher transmission. A PAL consists of a number of small prisms arranged in columns perpendicular to the optical axis. The column height decreases along the optical axis so that the projection of lens material is approximately linear with a Fresnel phase-plate pattern superimposed on it. The focusing effect is one dimensional, and the lens is chromatic. Hence, unwanted energies can be blocked by placing a slit in the image plane of a desired energy. We present the first experimental and theoretical results on an energy filter based on a silicon PAL. The study includes an evaluation of the spectral shaping properties of the filter as well as a quantification of the achievable increase in dose efficiency compared to standard methods. Previously, PAL's have been investigated with synchrotron radiation, but in this study a medical imaging setup, based on a regular x-ray tube, is considered.

**Keywords:** x-ray imaging; energy filtering; x-ray optics; prism-array lens; mammography

## 1. INTRODUCTION

It is well known that the absorbed dose necessary to obtain an x-ray image with sufficient signal-to-noise ratio varies with object thickness and x-ray energy spectrum.[1–3] In fact, for a certain object thickness there exists an optimal energy; lower energy photons will to a large extent be absorbed and contribute to dose, whereas photons of higher energies will pass the object merely unaffected and add to noise without giving much useful information. A spectrum that is narrow and centered around the optimal energy is hence the most dose efficient one.

Today, absorption filtering is the dominant method to narrow the x-ray tube spectrum. A thin film of material is used to filter out low energy photons, often in combination with a limited x-ray tube acceleration voltage in order to cut off higher energy photons.[4] In principle, the spectrum can be made very narrow by squeezing the maximum energy, set by the acceleration voltage, towards a wall of heavy filtration, however, only at the cost of a severe reduction in flux. The material of the absorption filter can also be chosen so as to have an absorption edge above the optimal energy to further reduce the high energy part of the spectrum. In general, however, the method of absorption filtering has not changed much since Pfahler, at the advent of x-ray imaging, discovered that putting a piece of leather between the patient and the x-ray tube resulted in less irritation on the skin of the patient.[5]

Nevertheless, several methods have in recent years been proposed to optimize the x-ray spectrum beyond what is practically achievable with absorption filtering. Very efficient monochromators can be applied to high flux x-ray sources, for instance synchrotrons,[6–8] laser plasma sources,[9] channeling radiation sources,[10] parametric x-ray sources,[11] and sources producing x-rays by inverse Compton scattering on free electron lasers.[12] Although

---

Electronic mail: fberg@particle.kth.se

image quality and dose efficiency are improved, the high complexity and cost of such sources limit the feasibility for routine clinical x-ray imaging.

Therefore it has also been proposed to apply less stringent energy filtering to a regular x-ray tube, resulting in a broader spectrum but also a better photon economy. Mosaic crystals, for instance, have small imperfections in the crystal structure and can be employed to achieve a spectrum optimized for mammography.[13–15] Another approach is to use chromatic refractive x-ray optics, *e.g.* a multi-prism lens (MPL), in a filtering set-up.[16–18]

An MPL consists of two arrays of prisms put on an angle in relation to each other. Peripheral rays entering the lens will encounter a larger number of prisms than will central ones, hence experiencing a greater refraction. Since the refractive index of the lens material varies with x-ray energy the lens is chromatic and unwanted energies can be blocked by placing a slit in the image plane of the desired energy. The problem with MPL filters is the high absorption, in particular at the periphery of the lens, which limits the usable aperture.

The prism-array lens (PAL) is a further elaboration of the MPL, with material corresponding to a phase shift of integer steps of $2\pi$ removed, resulting in a greater aperture, higher transmission, and shorter lens. PAL's are thus similar to Fresnel phase-plates, but with smaller aspect ratios, which facilitates manufacturing. In the past, PAL's have been investigated at synchrotron facilities,[19,20] but not with regular x-ray tubes in the context of medical x-ray imaging.

In the following, an investigation of the PAL for medical x-ray imaging will be presented. Experimental measurements of the focusing and filtering capabilities of the PAL will be compared to theoretical estimates obtained by diffraction and ray-tracing models. Furthermore, the feasibility of introducing the obtained spectra in a medical imaging set-up for dose reduction or increased signal-to-noise ratio will be evaluated, theoretically as well as experimentally.

## 2. THEORETICAL BACKGROUND

### 2.1. Prism-array lenses

Prism arrays can be set up in several ways, and for this study a design according to Figs. 1b) and c) was chosen. It consists of a number of small prisms arranged in columns in the $y$-direction (the height direction of the lens, perpendicular to the optical axis), which are displaced in the $y$-direction with an increasing displacement along the $x$-direction (the direction of the optical axis). The design is altered from the one presented in Ref. 19 by converting the regular prisms into right-angled half-prisms, and the sparser arrangement was found to be favorable for manufacturing due to fewer narrow corners and acute angles. Support structures were added at the entrance of the lens and for each prism column.

In analogy with Ref. 19, the projected amount of lens material in the $x$-direction ($X$) as a function of the distance from the optical axis ($y_1$) for a lens with $N$ columns can be described as

$$X(y_1) = 2T + \sum_{j=0}^{N} x_j(y_1) \qquad x_j(y_1) = \begin{cases} \mathrm{mod}\left(\frac{|y_1| - d(0.5+j)}{\tan\theta}, b\right) + t & y_1 \geq d(0.5+j) \\ t & y_1 < d(0.5+j) \end{cases}, \qquad (1)$$

where $x_j(y_1)$ is the projection in the $j$:th column. $T$ and $t$ are the support structure thicknesses at the entrance of the lens and at each column respectively. The columnar displacement is denoted $d$, and $b$ and $\theta$ are the prism base and angle. The modulus function represents the remainder after division. To minimize absorption, $b$ is chosen so as to correspond to a phase shift of a low multiple, ideally 1, of $2\pi$. The projection is approximately linear with a Fresnel phase-plate pattern superimposed on it, which results in a focusing effect in one dimension. In other words, peripheral rays encounter a larger number of prisms on their way through the lens than do central ones, thus also experience a greater refraction. The projected phase-plate pattern is an approximation built up of straight line segments, which is better the smaller $d$ is, given a certain prism size.

Imaging with a PAL follows for a thin lens the gaussian lens formula, $F^{-1} = s_o^{-1} + s_i^{-1}$, where $s_o$ and $s_i$ are the object and image distances respectively. The focal length is $F = d\tan\theta/\delta$, where $\theta$ is the prism angle, and $\delta$ is the decrement of the real part of the refractive index from unity. At energies and for lens materials of interest in this study, $\delta$ varies approximately as $E^{-2}$, and therefore the lens is chromatic, thus having different focal lengths for different x-ray energies. By placing a slit at the image plane ($s_i$) of a particular x-ray energy ($E_{\text{peak}}$) that very energy is transmitted, whereas other energies are out of focus and will be preferentially blocked, see Fig. 1a).

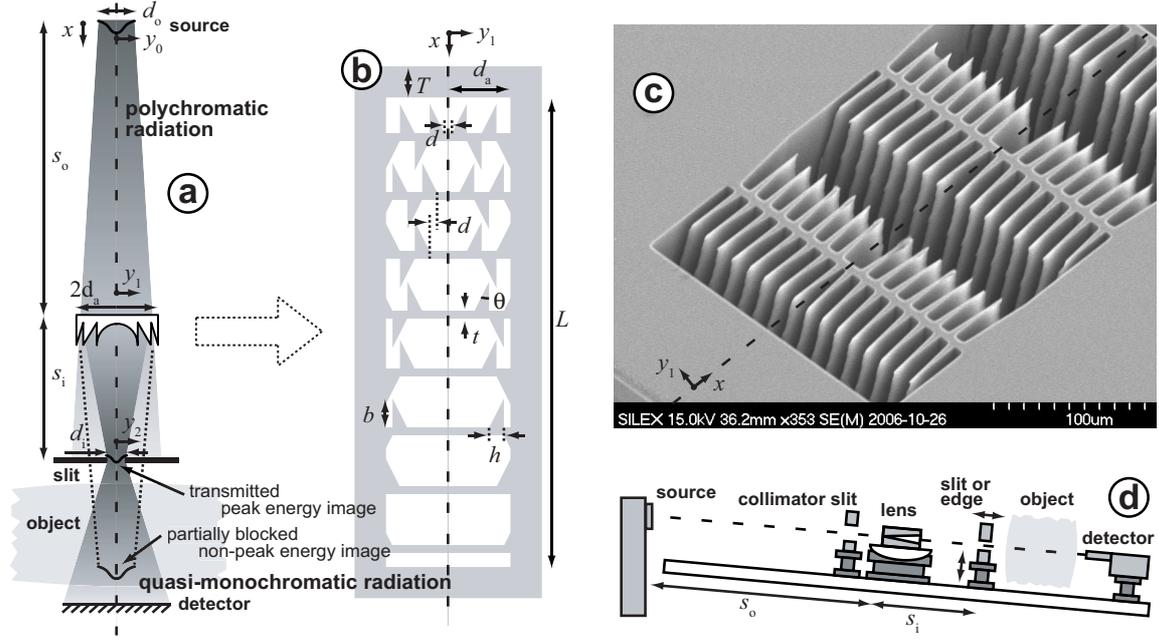

**Figure 1.** a) Energy filtering and imaging with the PAL. Vertical and horizontal axes are not to scale. b) A close-up of the PAL with lens parameters indicated. The lens used for this study had a total of 960 prisms, each with a base width $b = 59$ $\mu$m, a height $h = 6$ $\mu$m, and a prism angle $\theta = 5.8°$. c) SEM picture of the PAL used for the experiment. The optical axis is indicated by the dashed line. d) Schematic of the experimental set-up. The two 50 $\mu$m slits which collimate the beam in the $z$-direction are not shown.

## 2.2. Phase-plate and ray-tracing models

In a thin lens approximation, a PAL can be regarded as a one-dimensionally focusing Fresnel phase-plate, and its refractive effect can be described with Kirchhoff's scalar diffraction theory (see for instance Ref. 21 for a treatment of the subject). In one dimension, the time independent optical disturbance $\psi$ of a wave of length $\lambda$ and number $k = 2\pi/\lambda$ at $y_2$ in the image plane can be found,

$$\psi(y_0, y_2) = -i\sqrt{\frac{\varepsilon_0}{\lambda}} \int_{d_a} \frac{\exp\left[ik(\rho + r + nX)\right]}{\sqrt{\rho r}} K \, dy_1, \quad (2)$$

where the integral is taken over the lens aperture. $\varepsilon_0$ is the source strength per unit length, $\rho$ is the distance from $y_0$ at the source to $y_1$ at the lens, and $r$ is the distance from $y_1$ to $y_2$. Phase shift and absorption are described by the real ($\delta$) and complex ($\beta$) parts of the index of refraction, $n = 1 - \delta + i\beta$. $K$, finally, is the obliquity factor, which depends on the angle between vectors $\bar{\rho}$ and $\bar{r}$ at $y_1$ and is unitary in the forward direction. In the PAL set-up presented here, distances in the $y$-direction are very small compared to distances in the $x$-direction and $K$ is approximately unity. Furthermore, $\rho$ and $r$ in the denominator of Eq. (2) can be approximated with $s_o$ and $s_i$. The same approximation is not fair for the exponential due to a large $k$, but binomial expansions give

$$r = \sqrt{s_i^2 + (y_1 - y_2)^2} \approx s_i + (y_1 - y_2)^2/2s_i \quad \text{and} \quad \rho = \sqrt{s_o^2 + (y_1 - y_0)^2} \approx s_o + (y_1 - y_0)^2/2s_o, \quad (3)$$

which leads to

$$\psi(y_0, y_2) = -\frac{\sqrt{\varepsilon_0} i \exp\left[ik(1 + s_\mathrm{o} + s_\mathrm{i})\right]}{\sqrt{\lambda s_\mathrm{o} s_\mathrm{i}}} \int_{d_\mathrm{a}} \exp\left[ik\left(\frac{(y_1 - y_0)^2}{2s_\mathrm{o}} + \frac{(y_1 - y_2)^2}{2s_\mathrm{i}} + (i\beta - \delta)X\right)\right] \mathrm{d}y_1. \quad (4)$$

The intensity in the image plane ($\Phi_\mathrm{foc}$) over an interval corresponding to the image size ($d_\mathrm{i}$) when focusing radiation from a source of size $d_\mathrm{o}$ is thus

$$\Phi_\mathrm{foc} = \int_{d_\mathrm{o}}\int_{d_\mathrm{i}} |\psi(y_0, y_2)|^2 \, \mathrm{d}y_0 \, \mathrm{d}y_2. \quad (5)$$

In a similar set-up without a lens, the intensity from divergent radiation ($\Phi_\mathrm{div}$) over the same interval is

$$\Phi_\mathrm{div} = \frac{d_\mathrm{i}}{s_\mathrm{o} + s_\mathrm{i}} \int_{d_\mathrm{o}} \varepsilon_0 \, \mathrm{d}y_0. \quad (6)$$

The fraction of Eqs. (5) and (6) yields the gain of flux behind a slit of size $d_\mathrm{i}$ in the image plane compared to a set-up without lens,

$$G(E) = \frac{\Phi_\mathrm{foc}}{\Phi_\mathrm{div}}. \quad (7)$$

The average transmission of the PAL ($N_\mathrm{t}$) can be used to estimate the efficiency of the PAL filter, and is obtained by integrating and averaging the squared absorption part of Eq. (4) over the lens aperture,

$$N_\mathrm{t} = \int_{d_\mathrm{a}} \frac{\exp(-2k\beta X)}{d_\mathrm{a}} \, \mathrm{d}y_1, \quad (8)$$

where $-2k\beta$ corresponds to the linear attenuation coefficient of the lens material.

The described phase-plate model assumes the lens to be thin compared to other distances in the $x$-direction, but the length of the lens is at least $L = Nb$, which is typically in the order of a centimeter. To study the effect of such approximations a full ray-tracing model was set up, and the law of refraction was used to calculate the deflection angle in each tooth of the lens. Roughly $10^5$ rays were used, emerging from the source at random positions and angles, and with a flat energy spectrum. The ray-tracing model still suffers from the approximations that no scattering occurs, and that the lens is flawless.

## 3. MATERIAL AND METHODS

### 3.1. Experimental set-up

PAL's with various prism parameter settings were fabricated by a commercial company[*] using deep reactive ion etching (DRIE) of silicon according to the so-called Bosch process. It is a plasma-based cyclic process which can be used to attain micro-structures with aspect ratios of up to 40:1, and the etching process can be tuned for optimization of side-wall verticality and reduction of side-wall roughness. See Ref. 22 for a thorough description of the process. For masking, the surface of the silicon wafer was oxidized, and the thin layer of silicon dioxide so created was patterned using standard lithographic methods. The mask was removed with hydrofluoric acid after the etching.

For the present study, a lens was chosen from the batch, having a total of 960 prisms, each with a prism base $b = 59$ $\mu$m (which corresponds to a phase shift of $2\pi$ in silicon), height $h = 6$ $\mu$m, and prism angle $\theta = 5.8°$. See Fig. 1b). There were $2 \times 16$ prisms in the first column, which gives a physical aperture of $2d_\mathrm{a} = 192$ $\mu$m, and the displacement in the $y$-direction between two adjacent columns was $d = 1.6$ $\mu$m. These parameters optimize the lens for 23 keV with a focal length of 172 mm at that energy. Support structures were added at the entrance and exit of the lens ($T = 100$ $\mu$m) and at each column ($t = 8$ $\mu$m). The $N = 63$ columns yielded a lens length of 9 mm. In Fig. 2 the projection of lens material in the $x$-direction according to Eq. 1 is shown. The manufacturing process was not totally optimized and the lenses exhibited structure failure, in particular over-etch of the convex prism angles of up to 10 $\mu$m, which is to be seen in Fig. 1c). This limited the depth of the lens to 165 $\mu$m.

---

[*]Silex Microsystems (Järfälla, Sweden)

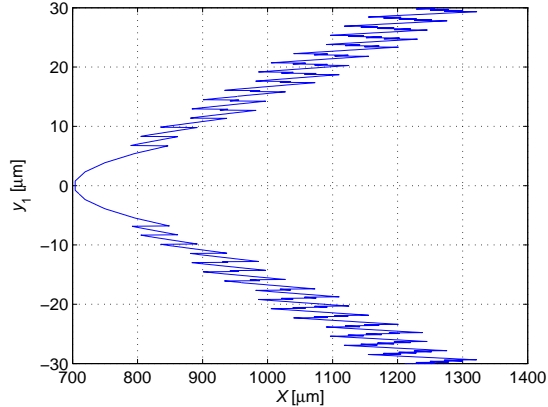

**Figure 2.** Projected lens profile ($X$) as a function of the distance from the optical axis ($y_1$) for the PAL used in the study. Only a part of the aperture is shown.

A tungsten anode x-ray tube,[†] acceleration voltage 33 kVp and anode current 50 mA, was employed in a set-up depicted in Fig. 1d). A CZT compound solid state detector[‡] with a near 100% detection efficiency and negligible hole tailing around 23 keV (according to the manufacturer) was used. An energy resolution of 0.5 keV was chosen for sufficient count rate without pile-up. To facilitate line-up, the lens was mounted on precision stages for lateral translation and tilting. Upstream of the lens a 200 $\mu$m collimator slit was placed to match the aperture of the lens and to reduce stray radiation. The beam was further collimated in the $z$-direction (the depth direction of the lens, perpendicular to the optical axis) to 50 $\mu$m by two slits; one in front of the lens and one in front of the detector.

At the image plane ($s_i$), a precision stage for translation in the $y$-direction was placed, on which could be mounted either a 12 $\mu$m tungsten slit or a sharp tantalum edge depending on the measurement (see below).

In order to determine the x-ray tube focal spot size in the $y$-direction, the anode angle was measured and the source size could be determined from specifications by the manufacturer. This source size was used for the models. For verification, an edge scan was performed with the tantalum edge in the image plane against the upper edge of the collimator slit prior to the experiments. The focal spot size in the $z$-direction was 12 mm, and could be considered unlimited for the present measurements.

The distance $s_o = 585$ mm was chosen in accordance with the gaussian lens formula so as to produce a sharp 23 keV image of the source at $s_i = 244$ mm. These distances refer to the entrance of the lens, which is assumed to be the plane of the lens, but the lens is in practice not thin and a somewhat different image distance can be expected from experiment and ray-tracing, even for a perfect lens. Therefore, the stage at $s_i$ could be moved in the $x$-direction, and measurements were performed in steps along the optical axis in order to find the image plane for 23 keV radiation. Image sizes predicted by the models were recalculated for the measured image distance in order to make comparisons meaningful. This was also done for the predicted gain, which varies with image size and distance.

### 3.2. Focusing and filtering properties

To obtain a precise measurement of the source size and of the gain, the tantalum edge was scanned in steps of 1 $\mu$m over the image plane. The data so obtained correspond to the integral of the intensity profile in the image plane as a function of position, and so the actual profile was found by differentiation. A gauss function was fitted to the differentiated edge scan profile, and the image size ($d_i$) defined here as the full-width-at-half-maximum (FWHM), was calculated.

---

[†]Philips PW2274/20 with high tension generator PW1830
[‡]Amptek XR-100T-CZT

To obtain $\Phi_{\text{foc}}$, a straight 12 $\mu$m line was fitted to the edge scan profile over the interval with the fastest rate of change, which is where the edge passes the image of the source. The slope of the line was taken to be the derivative, which is a good measure of the flux over an interval where the scan profile is essentially linear, and the length of the straight line corresponds to the size of the slit. A second edge scan without the lens was performed in order to determine $\Phi_{\text{div}}$. Since in this case the rate of change was much lower, the straight line was fitted over 50 $\mu$m to reduce statistical noise, and normalized to 12 $\mu$m. An alternative way to calculate the gain would have been to use the fitted gaussian profile, but such a calculation is more dependent on the actual source shape.

The average transmission factor of the lens ($N_t$) was derived from the edge scan as the fraction of the two non-differentiated profiles at a point where there was no shadow of the edge.

### 3.3. A model system for medical imaging

Although an edge scan avoids some of the problems compared to measurements with a slit, *e.g.* alignment and change of slit size, an authentic filtering set-up does require a real slit in the image plane. Therefore such measurements were also performed with a $\sim$ 12 $\mu$m tungsten slit.

The spectral quantum efficiency for a certain mammographic spectrum and breast size can be defined as[23]

$$\text{SQE} = \frac{\text{SDNR}_s^2}{\text{AGD}_s} \cdot \frac{\text{AGD}_m}{\text{SDNR}_m^2}. \tag{9}$$

Here, $\text{SDNR}_s$ and $\text{AGD}_s$ are the signal-difference-to-noise-ratio and average glandular dose for the investigated spectrum. $\text{SDNR}_m$ and $\text{AGD}_m$ are the same quantities for the ideal monochromatic case. For a quantum limited system, the SQE is an exposure independent quantity, inversely proportional to the dose needed to obtain a certain SDNR.

Dose efficiency of the PAL filter was experimentally evaluated using the SQE metric with a simple mammography phantom. The phantom consisted of 6 cm tissue equivalent material (BR12), corresponding to a breast of approximately 50% glandularity, with an added 200 $\mu$m aluminum foil, representing the contrast of a micro calcification. Values of the SDNR for the phantom were experimentally obtained for PAL and aluminum filtered spectra according to

$$\text{SDNR} = \frac{|n_1 - n_2|}{\sqrt{n_1 + n_2}}, \tag{10}$$

where $n_1$ and $n_2$ are the signals from pixels with and without aluminum foil. To find the AGD, monte-carlo calculated normalized glandular dose coefficients[24] were applied to the spectra. The ideal monochromatic SDNR was obtained theoretically with material compositions,[24] and x-ray attenuation coefficients.[25] For comparison, values of the SQE were also calculated for the experimentally obtained spectra at a range of breast thicknesses, and the ray-tracing gain was applied directly to the tungsten spectrum to estimate the theoretically highest possible SQE.

The impact of the PAL filter on spatial resolution and image acquisition time in medical imaging will depend on the imaging geometry and is only briefly discussed in this paper.

## 4. RESULTS

At the 3.5($\pm$0.2)° anode angle being used, the x-ray tube focal spot size was specified by the manufacturer to 24.5($\pm$1.5)$\mu$m. The edge scan yielded a 24.5(21.5 − 28.5)$\mu$m FWHM fitted gaussian profile, and thus confirmed the source size.

The image plane of the set-up was found on a distance $s_i = 301$ mm for 23 keV radiation, 57 mm farther from the lens than expected.

Fig. 3 shows the result of an edge scan at the image plane. The energy is 23 keV and the 12 $\mu$m fitted line used to find the peak gain is indicated. In Table 1 the average transmission factors ($N_t$) of models and experiment are tabulated and can be compared to Eq. (8) which yields 0.35. Also shown are the FWHM's of the fitted profiles (*i.e.* $d_i$). Values of $N_t$ and $d_i$ derived from the models are recalculated for $s_i = 301$ mm.

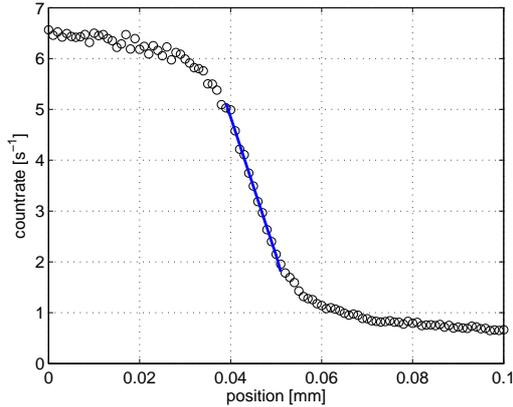

**Figure 3.** Edge scan at 23 keV of the PAL filtered 33 kVp spectrum in the image plane of the set-up; count rate as a function of position of the edge. The fitted 12 $\mu$m line is indicated.

**Table 1.** Results from the filtering and focusing measurements at 23 keV, compared to the phase-plate and ray-tracing models for a 24.5 $\mu$m FWHM gaussian source. The image size ($d_i$) and peak gain ($G$) of the models are recalculated for the image distance ($s_i$) of the measurement. Also presented is the average transmission factor ($N_t$).

|                   | $s_i$ [mm] | $d_i$ [$\mu$m] | $G$ | $N_t$ |
|-------------------|------|-------|-----|------|
| Measurement       | 301  | 17.3  | 5.2 | 0.32 |
| Phase-plate model | 236  | 13.5  | 6.7 | 0.34 |
| Ray-tracing model | 246  | 12.4  | 6.5 | 0.31 |

Relative flux after a 12 $\mu$m slit as a function of energy is plotted in Fig. 4 for filtered and unfiltered spectra. The filtered spectrum has an FWHM of 4.7. These curves yield the gain by division, with the result shown in Fig. 5 along with the predicted results from the phase-plate model as well as from ray-tracing. Also shown in Fig. 5 is ray-tracing gain for 21.5 and 28.5 $\mu$m sources, which correspond to $\pm 1$ standard deviation of the source size measurement. Values of the peak gain are tabulated in Table 1, where the gain of the models are recalculated for $s_i = 301$ mm.

In Fig. 6 is plotted the experimentally obtained PAL and aluminum filtered spectra, and the unfiltered tungsten spectrum multiplied with the ray-tracing gain. The unfiltered spectrum is also shown. Values of the FWHM for the filtered spectra are presented in Table 2.

The SQE values of PAL and 0.5 mm aluminum filtered spectra for the 6 cm breast phantom are indicated in Fig. 7. Also presented is the calculated SQE as a function of breast thickness for the experimentally obtained PAL and aluminum spectra, and for a PAL spectrum obtained from the ray-tracing model. The peak values of the SQE are tabulated in Table 2 along with the dose reduction of the PAL compared to aluminum filtered spectra.

## 5. DISCUSSION

### 5.1. Focusing and filtering properties

The measured image of the source is larger than would be expected from the models, and corresponds to a source size of 33.5 $\mu$m. This is not covered within one standard deviation of the source size measurements, but it was found that the source size varied with the cooling water temperature, which could not be kept totally constant. The measurements were long compared to the variations, and a moving source will therefore show up as a larger and blurred image. Moreover, imperfections in the lens structure might have influenced the result. As can be seen in Fig. 1c), the over-etch causes the prism size to vary along the $z$-direction, which means that also the focusing properties might vary with depth. A narrow beam of only 50 $\mu$m, which is about one third of the lens depth, was used to obtain uniform focusing, but still the effects of varying prism sizes cannot be excluded.

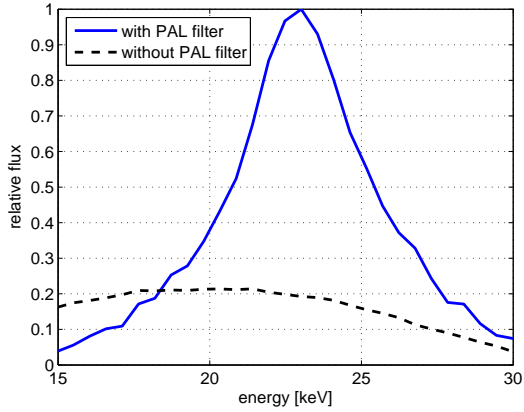

**Figure 4.** Relative flux after a 12 $\mu$m slit as a function of energy for the PAL filtered 33 kVp spectrum and for the unfiltered spectrum (*i.e.* $\Phi_{\text{foc}}$ and $\Phi_{\text{div}}$). The FWHM of $\Phi_{\text{foc}}$ is 4.7 keV.

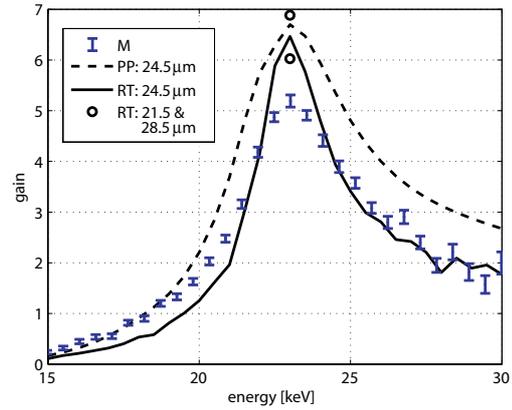

**Figure 5.** Measured gain (M) as a function of energy compared to the phase-plate (PP) and ray-tracing (RT) models for a 24.5 $\mu$m source. Also indicated is ray-tracing gain for source sizes 21.5 and 28.5 $\mu$m, which correspond to $\pm 1$ standard deviation of the source size measurement. Error bars correspond to $\pm 1$ standard deviation of the measured gain.

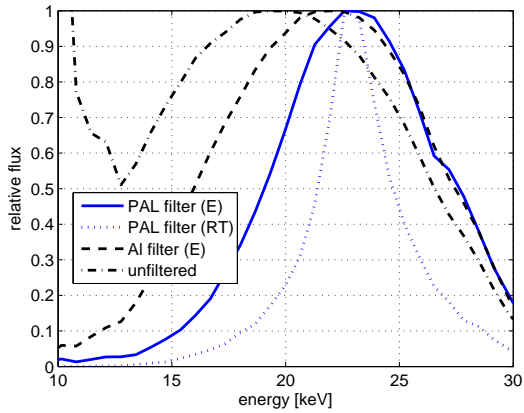

**Figure 6.** The experimentally obtained PAL and aluminum filtered spectra (E), and the raw tungsten spectrum multiplied with the ray-tracing gain (RT). The latter one corresponds to the best possible filtration with the current PAL.

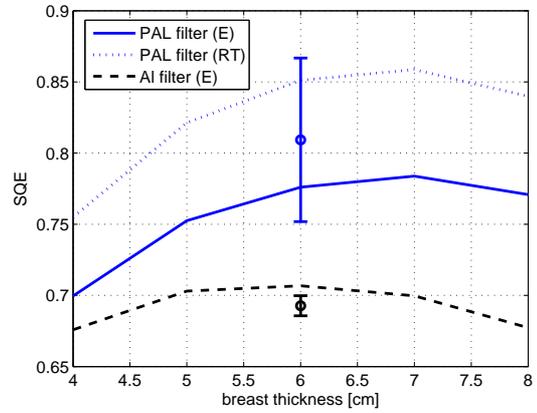

**Figure 7.** Experimentally obtained SQE of PAL and 0.5 mm aluminum filtered spectra for the breast phantom (E). Also shown is the calculated SQE as a function of breast thickness for the experimentally obtained PAL and aluminum spectra, and for a PAL spectrum obtained from the ray-tracing model (RT).

**Table 2.** Results from the measurements of dose efficiency. The SQE and dose reduction refer to 6 cm breasts. The experimentally obtained values (E) of the PAL and aluminum SQE are shown, along with the calculated SQE from the ray-tracing model (RT). Also presented are the FWHM's of the filtered spectra.

|  | FWHM [keV] | SQE | Dose reduction [%] |
|---|---|---|---|
| PAL (E) | 8.6 | $0.81 \pm 0.06$ | $14 \pm 6$ |
| PAL (RT) | 3.3 | 0.85 | 17 |
| Aluminum (E) | 11.5 | 0.69 | - |

Measured gain as a function of energy is in good agreement with the ray-tracing model at all energies. The deviation at the peak energy of 20% can again be explained by a larger than expected or moving source, which will result in a lower gain. The effect is illustrated in Fig. 5 by the ray-tracing gain for source sizes 21.5 and 28.5 $\mu$m. Imperfections in the lens structure might influence also the gain, by non-perfect focusing or absorption at under etched narrow passages. It should also be noted that the slit size of 12 $\mu$m was kept constant, although if adapted to the size of the image would yield a higher gain. The reason was to make comparisons with the models, and with the results using the non-adjustable tungsten slit, more lucid.

Scattered and reflected rays are treated as absorption in the ray-tracing model. Small angle Rayleigh scattering might, however, account for parts of the increased image size, whereas Compton scattering and reflection should add background radiation and reduce the gain. In fact, the edge scan in Fig. 3 does indicate a certain amount of background radiation since it has a superimposed slow rate of change and does not reach zero after the image. The scattering cross section increases fast with energy at the considered energy interval, whereas total external reflection on the prism surfaces is more probable at low energies.

As can be seen in Fig. 5, there is good agreement between the phase-plate and ray-tracing models for predicted gain at the peak energy 23 keV, but the phase-plate gain decreases slower when moving to other energies, in particular to higher ones. The faster decrease towards lower energies is likely to be due to absorption in the support structures of the lens. This seems to indicate that the focal length predicted by the phase-plate model varies slower with energy than $E^2$, and since ray-tracing and measurement agree better, the reliability of the phase-plate model can be questioned. Since the lens length was almost 10 mm, the thin lens approximation might have influenced the result; the lens profile shown in Fig. 2 assumes a thin lens, but deflection will occur all the way through the PAL and in particular peripheral rays with a large deflection might miss prisms towards the exit side.

Moreover, the thin lens approximation can explain the difference in focal length between the two models, as seen in Table 1, since the phase-plate model assumes a thin lens but the ray-tracing does not. The measured elongation of the focal length, on the other hand, is larger than predicted by ray-tracing and much larger than the length of the lens. Imperfections in the lens structure is again a probable reason for the deviation.

### 5.2. Dose reduction in medical imaging

According to Fig. 7 and Table 2, the PAL filter can reduce the dose in mammography compared to absorption filtering with 14% for a 6 cm breast. The measured dose reduction does, however, suffer from fairly large statistical errors, and a more moderate approach would be to compare the SQE calculated from the measured spectra. In that case a reduction in dose of 9% can be expected. An upper bound is set to 17% by the ray-tracing calculation using only theoretical values for PAL and aluminum filtered spectra. As can also be seen in Fig. 7, the PAL filtered spectrum is better optimized for 7 cm breasts for which a larger dose reduction can be expected.

Attenuation coefficients of real breast tissue and not BR12 was used to calculate the SQE, which might explain the small deviation of the measured SQE from theoretical values for the aluminum filtered spectrum.

As can be seen in Table 2 and Fig. 6, the PAL filtered spectrum is substantially broader than the spectrum predicted by ray-tracing, which lowers the SQE. It is also broader than the spectrum obtained by the edge scan procedure (Fig. 4). Therefore, except for the sources of error discussed above, it is fair to assume that the 12 $\mu$m slit is broadening the spectrum. In fact, the thin walls of the slit were found to leak radiation, especially at higher energies, and that would indeed contribute to the broadening. Additionally, the size of the slit was associated with some uncertainty.

When interpreting the results it should also be kept in mind that the steep angle of the set-up leads to more self-filtration and a somewhat harder spectrum than in diagnostic radiology, where an angle of about 17° is common.[4]

### 5.3. Photon economy

The support structures of the PAL design presented here were added to allow for through etch of the silicon wafer, but the etch depth was, as explained above, limited by structure failure. By removing the support structures, the average transmission at 23 keV could have been increased to 0.57 according to Eq. (8), which is an improvement in the transmission and the gain by a factor 1.6.

Moreover, the atomic number, $Z$, of the lens material plays an important role for the transmission. The reason for choosing silicon in this study was that manufacturing methods are readily available and the transmission is good enough for a proof of principle study. For improving the lens, however, a lower $Z$ material, such as a plastic, would be preferable. If going from silicon ($Z = 14$) to epoxy ($Z \approx 6$) and increasing $b$ so that $\theta = 2.9°$ in order to keep the phase shift of $2\pi$, focal length, and aperture, the average transmission would become 0.93 for a lens with no support structures. This is an improvement of an additional factor 1.6.

Compared to its predecessor, the MPL, a PAL makes possible higher transmission since lens material corresponding to a phase shift of $2\pi$, inactive for deflection but active for absorption, is removed. In a previous experimental study,[17] the average transmission of an epoxy MPL was found to be higher than the present PAL transmission (0.5 at 23 keV), but the MPL was made of epoxy and the result should be well within reach of an improved PAL. Additionally, the gaussian transmission of the MPL approaches zero faster towards the periphery of the lens than does the approximately linear transmission profile of the PAL, which opens the possibility for a larger aperture and a higher gain. In fact, an aperture improvement factor (AIF) proportional to the improvement in gain, was derived in Ref. 19 as AIF $= 3.2 \cdot \sqrt{\delta F} \cdot (\sqrt{\mu} b \tan \theta)^{-1}$. For epoxy lenses at 23 keV the improvement is a factor 23.

To be feasible for medical imaging, the PAL filter must allow a photon flux and image acquisition time comparable to existing imaging systems. Since the lens provides a line focus which is not limited in length, it may be coupled to a row detector in a scanning system, such as computed tomography or scanned slit mammography systems. Furthermore, several lenses in an array can be coupled to a row detector each in a multi slit geometry. The aperture of the PAL is approximately the same size as a collimator slit in a conventional such geometry, and the transmission of the lens may therefore be compared to the transmission of an absorption filter. For instance, the transmission of a 0.5 mm aluminum filter at 23 keV is 0.73, which is roughly a factor 2 higher than the transmission of the lens presented here, but within reach of an improved PAL with less support structures and a more optimized lens material. Additionally, since the PAL gathers radiation, the lens aperture can be made larger than the collimating slit in a conventional system at the same resolution. The actual flux integrated over all energies does, however, depend on several specific parameters of the set-up including its length, the resolution, and the size of the source, and a more extensive study would be needed to address the issue of image acquisition time properly.

### 6. CONCLUSIONS

An energy filter for medical x-ray imaging based on a silicon prism-array lens with a peak energy at 23 keV has been investigated, experimentally and theoretically.

When imaging a bremsstrahlung source, the lens produces a line shaped image with a size somewhat larger than expected. Measured gain of flux is in good agreement with the ray-tracing and phase-plate models at the peak energy, deviating -20% from ray-tracing. Deviations in both image size and gain can be accounted for by instability of the x-ray tube focal spot size and position, imperfections in the lens structure, and approximations in the models. For energies other than the peak energy the gain predicted by the ray-tracing model agrees well with measurements, whereas the phase-plate model predicts less efficient filtering. This is likely due to approximations in the model.

An experimental model of a mammography system showed that the PAL filter can reduce dose by 14% compared to absorption filtering for 6 cm breasts. The dose reduction in this study was limited by imperfections in the set-up, and potentially a dose reduction of 17% is possible according to ray-tracing.

Change in lens design can increase the transmission and gain by a factor $1.6 \times 1.6$, and the aperture of the PAL can be increased with more than an order of magnitude compared to its predecessor, the MPL. It is estimated that a filter based on such an improved lens will allow a photon flux comparable to conventional absorption filters.

## ACKNOWLEDGMENTS

We acknowledge the Swedish Research Council and and the foundation Lars Hiertas Minne for funding parts of our work on prism-array lenses.